# Software Engineering Knowledge Areas in Startup Companies: a mapping study

Eriks Klotins, Michael Unterkalmsteiner, Tony Gorschek

Blekinge Institute of Technology, SE-37179, Karlskrona, Sweden
{eriks.klotins, michael.unterkalmsteiner, tony.gorschek}@bth.se

**Abstract.** *Background* – Startup companies are becoming important suppliers of innovative and software intensive products. The failure rate among startups is high due to lack of resources, immaturity, multiple influences and dynamic technologies. However, software product engineering is the core activity in startups, therefore inadequacies in applied engineering practices might be a significant contributing factor for high failure rates. *Aim* – This study identifies and categorizes software engineering knowledge areas utilized in startups to map out the state-of-art, identifying gaps for further research. *Method* – We perform a systematic literature mapping study, applying snowball sampling to identify relevant primary studies. *Results* – We have identified 54 practices from 14 studies. Although 11 of 15 main knowledge areas from SWEBOK are covered, a large part of categories is not. *Conclusions* – Existing research does not provide reliable support for software engineering in any phase of a startup life cycle. Transfer of results to other startups is difficult due to low rigor in current studies.

## 1 Introduction

Recent developments in technologies have created an increasing demand for innovative software products. Startup companies are addressing this need and gain importance as suppliers of software-intensive products and innovation. The inherent nature of software enables small companies to produce and launch software products fast with few resources. However, most of startup companies fail before realizing any significant achievements [11]. Partially this is due to market factors or financial issues, however the impact of software product engineering and inadequacies in applied engineering practices is not fully explored, and might be a significant contributing factor for the high failure rates.

Chorev et al. [8] identify 16 key factors for a successful startup, such as political and economical environment, marketing, idea, funding and product development among others. Many authors [2, 3, 8, 12, 26, 41] address general issues of startups. Only a few focus on how software engineering is done in startups. Yau et al. argue that scaled down engineering practices solve problems present in larger, established companies while ignoring specific challenges that emerge only in startup companies, stating that different approaches altogether are needed for software engineering in the context of startups [20].

In this paper we aim at identifying software-intensive product engineering practices utilized in startup companies and mapping them to Software Engineering Body of Knowledge (SWEBOK) [31] knowledge areas and categories, describing both state-of-the art, and gaps in research on startup software engineering. Furthermore, to analyze how identified software engineering knowledge areas support the startup life cycle we use the four phase model proposed by Crowne [11] and map identified knowledge areas to different phases in the startup life-cycle. By use of these well-established taxonomies [2], [10] we show state-of-the-art and expose gaps for further research, but with a clear and distinct focus on the software engineering perspective.

This paper is structured as follows. Section 2 gives an overview of the field and motivates the study. Section 3 details the research methodology we applied to identify and map relevant papers. Section 4 reports results from the mapping. Section 5 answers the research questions and discusses the results. Section 6 concludes the paper.

## 2 Background and related work

A startup company shares many features with small or medium enterprises such as youth, market pressure and dynamic technologies [33]. However startups are different due to their aim and the challenges they face [33]. In contrast to established companies, who regardless of their size focus on optimizing an existing business model, startups focus of finding one [26]. Sutton [33] defines a startup as an organization that is challenged by youth and immaturity, extremely limited resources, multiple influences and dynamic technologies and markets.

Crowne [11] had proposed a four phase start-up life-cycle model. Successfully transferring from first phase to the last indicates that a startup has become an established company. The model identifies distinct challenges at each phase that a start-up must address to advance to the next stage. We seek to identify knowledge areas supporting transfer trough start-up life cycle by addressing challenges identified by Crowne [11].

Paternoster et al. [23] conducted a mapping study to characterize state-of the-art research in startups. They conclude that only a minority of studies in the area are dedicated to (software) engineering, and since 2000 when this gap was first identified [33] it has been only partially filled.

Coleman et al. [9] conducted a grounded theory study to explore how software processes are formed in a startup. This study concludes that there is not enough resources to explore the best way to develop the software and startups use whatever software process that supports their immediate business objective. Consequently, the development process is heavily influenced by previous experiences of a person acting as development manager [9].

Pino et al. [25] conducted a systematic review on software process improvement (SPI) in small and medium organizations. The study is aimed at discovering what approaches to SPI in small-medium companies exist. Although their study was not aimed at startup organizations, they conclude that prescriptive approaches, such as CMM and SPICE, are not suitable for small organizations. Therefore, they emphasize the need for more lightweight and tailored approaches.

Several startup specific process models have addressed this need. For example, LIPE [40] addresses immaturity, ad-hoc approaches and scalability of engineering processes. ESSDM [4] proposes an iterative approach to build and validate multiple product ideas simultaneously. The Helical model [13] supports innovation by experimentation of multiple product ideas, frequent releases and synchronization with other organizational processes.

Software Engineering Body of Knowledge (SWEBOK) characterizes content of software engineering discipline and promotes consistent view to software engineering. SWEBOK is organized in 15 main knowledge areas; each knowledge area is organized in sub-categories. Although, SWEBOK is not specifically aimed at startups it is widely recognized within software engineering community [31].

To understand the degree to which research supports software engineering in startups, it is useful to map existing studies. One recent contribution is the mapping study by Paternoster et al. [23], describing research on startups and providing a characterization of software development in the startup context. However, their work does not classify the identified work practices such that it can be understood what software engineering problem is actually addressed. In contrast, our study aims at identifying and classifying software engineering knowledge areas in startup companies, enabling a) analysis and improvement of existing practices and b) revealing opportunities for further investigation.

## 3  Research methodology

The mapping process consists of three activities: identification of relevant publications, data extraction, and data mapping. We identify relevant publications by an emerging systematic literature review method – snowball sampling [38]. For data mapping we follow the recommendations by Petersen et al. [24].

### 3.1  Research questions

Our study is driven by the goal to understand to what extent engineering in startup companies is supported by research. To pursue this goal we seek answers to the following research questions:

**RQ1:** What is state-of-practice in terms of utilization of software engineering knowledge areas in startups?

**RQ2:** What is the relevance and rigor of the studies reporting experiences from software engineering in startups?

In order to structure the identified practices into knowledge areas, as well as identify gaps in knowledge (RQ1) we use SWEBOK [31] as a software engineering dictionary. Although SWEBOK was not created for startups, we lack alternatives, and SWEBOK is considered the accepted SE subject area overview [6, 28]. To provide an account whether the practices can be transferred to industry (RQ2) we assess rigor and relevance [17] of the identified studies.

## 3.2 Mapping study design overview

**Identification of primary studies:** We used snowball sampling [38], defining the starting set from an earlier and broader mapping study on startups [23]. We performed only forward snowball sampling from the starting set, as earlier papers are likely to be covered by the previous study by Paternoster et al. [23].

We screened the sampled papers to select studies that report on primary research focused on software engineering practices in startups. At first, for each paper we applied a sanity check filtering out duplicates, non-English and non-peer-reviewed papers. We used titles and abstracts for screening; in ambiguous cases, we read the full text. The screening criteria are summarized in table 1.

**Table 1.** Screening criteria

| Inclusion criteria | Notes | Examples of excluded papers |
|---|---|---|
| A paper reports primary research | With primary research we understand studies that provide direct evidence about the research question [16]. | [15, 34] |
| A paper reports a study in a startup company | We have used definition by Sutton [33] to differentiate between startups and established companies. | [22, 32] |
| A paper addresses software engineering | We use SWEBOK [31] to identify software engineering topics | [34, 37] |
| A paper addresses a challenge or a practice | With practice we identify use of a methodology, routine, tool or framework pertaining software engineering. With challenge we understand difficulty to achieve intended product quality, scope, budget or time constraints | [10] |

We used Google Scholar to identify referencing papers, i.e. to perform forward snowball sampling. The first author performed the screening of papers. Results of the process were organized in a spreadsheet that was reviewed by the second and third author.

**Data extraction:** Post identification of relevant studies data extraction was performed with the primary goal to extract information indicating which knowledge areas are explored in the study. We also extracted information pertaining to rigor – context description, description of study design, validity discussion, and relevance – information on subjects, study context, scale and research method according to the assessment method by Ivarsson et al. [17].

## 3.3 Analysis

To answer our first research question (RQ1: What is state-of-practice in terms of utilization of software engineering knowledge areas in startups?) we map the extracted practices to SWEBOK [31] knowledge areas and categories. In the mapping, we keep track on coverage – how many of knowledge areas and categories are covered by evidence. Coverage, or lack of it, reveals gaps in current research. We also use startup life cycle model by Crowne [11] to identify to what extent state-of-practice covers all four phases of startup life cycle.

To answer our second research question (RQ2: What is the relevance and rigor of the studies reporting experiences from software engineering in startups?) we synthesize rigor, relevance and research type, and analyze number of cases per study.

### 3.4 Threats to Validity

Systematic reviews have a generic bias towards positive results as they get published more often [5]. However, we do not consider this as a major threat as we especially aim to identify gaps and do not address the performance of individual practices. Another generic threat to mapping studies using snowball sampling is related to the quality of the starting set [38]. As a starting set we have selected the 43 studies identified by Paternoster et al. [23]. The set covers a rather broad period from 1994 to 2013, includes both journal and conference papers from multiple publishing venues. Thus, the starting set follows all guidelines set forth by Wohlin [38].

We focused on forward snowball sampling, as earlier studies are likely to be covered by the previous mapping study by Paternoster et al. [23]. Nevertheless, we performed a backward iteration on the final set of papers to reduce the risk of missing important studies. As a result, 241 papers were discovered. Subsequent screening identified one [20] relevant study. Furthermore, we have conducted a review of gray literature to screen further information pertaining to our research questions. This resulted in one more paper [12], which we did however not include in the further analysis because the described practices are already reported in other, peer-reviewed, studies.

Threats to study selection are addressed by explicit inclusion and exclusion criteria, and a detailed screening protocol. Explicit extraction templates guided the data extraction process, thus ensuring uniformity of the extracted data. To avoid bias set by personal opinions of the researchers executing the study, ambiguous cases were discussed among the authors.

## 4 Results

As a result of the snowball sampling, we identified 558 papers, 14 of them passed the screening process and were included for further analysis. The reasons for exclusion break down to the following: 80 duplicates, 17 not written in the English, 126 not peer reviewed (books, keynotes, blogs etc.), 354 not focused on startups, 50 not addressing software engineering, 7 not describing a practice or challenge, 32 not available in full text.

From the relevant papers we extracted 54 practices distributed among 11 of the 15 software engineering knowledge areas. Table 2 summarizes the identified primary studies and respective SWEBOK knowledge areas. The coverage column shows how many second level categories are covered by the papers (e.g. 6/8 means that two categories out of total of eight in SWEBOK were not covered at all).

**Table 2**. Knowledge areas and relevant papers

| Knowledge Area (KA) | Coverage | Covered categories |
|---|---|---|
| Software Requirements | 6/8 | Requirements Process [14]<br>Requirements Elicitation [1, 29]<br>Requirements Analysis [35]<br>Requirements Validation [1, 29]<br>Practical Considerations [19, 20] |
| Software Design | 4/8 | Software Design Fundamentals [1, 14, 29]<br>Key Issues in Software Design [18]<br>User Interface Design [1, 21, 30, 35]<br>Software Design Tools [1, 35] |
| Software Construction | 3/5 | Software Construction Fundamentals [7, 21, 29, 30, 36]<br>Managing Construction [7]<br>Practical Considerations [21] |
| Software Testing | 2/6 | Software Testing Fundamentals [18]<br>Test Process [19, 35] |
| Software Maintenance | 1/5 | Techniques for Maintenance [29] |
| Software Configuration Management | 3/7 | Software Configuration Identification [1]<br>Software Release Management and Delivery [1, 19, 29]<br>Software Configuration Management Tools [29] |
| Software Engineering Management | 3/7 | Software Project Planning [18, 29]<br>Software Project Enactment [39]<br>Software Engineering Management Tools [27] |
| Software Engineering Process | 2/5 | Software Process Measurement Techniques [20]<br>Software Engineering Process Tools [1] |
| Software Engineering Models and Methods | 2/4 | Modeling [1]<br>Software Engineering Methods [1, 13, 14, 21, 29] |
| Software Quality | 1/4 | Software Quality [18] |
| Software Engineering Professional Practice | 2/3 | Professionalism [1]<br>Communication Skills [1, 19, 21] |
| Software Engineering Economics | 0/5 | |
| Computing Foundations | 0/17 | |
| Mathematical Foundations | 0/11 | |
| Engineering Foundations | 0/7 | |

One of the main goals of research on startups is the transfer and widespread use of the results [17]. Potential for transfer can be judged by measuring rigor and relevance. The results reveal that most papers have high relevance, as they report studies performed in actual startups. However, the rigor of these papers is low as they lack contextual descriptions as well as in what manner the study was designed and executed. Figure 1 summarizes contribution type, rigor and relevance.

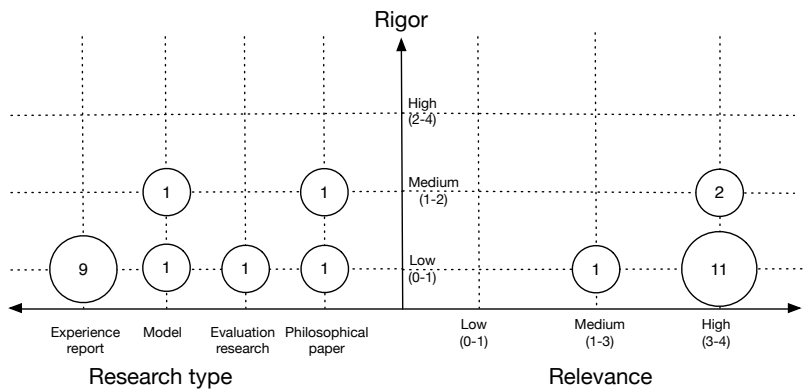

**Fig. 1.** Overview of research type, rigor and relevance distribution

As shown in figure 1, left side, the majority of the discovered papers are experience reports with low rigor, indicating a rather weak presentation of study design, industrial context and validity threats. The right side of figure 1 shows that the majority of the identified papers present results relevant for industry. The reported studies are conducted in a real industry environment, on a representative scale and are utilizing empirical research methods.

A study that investigates more than one case and compares findings among multiple cases provides more generalizability. We extracted the number of cases studied per paper and mapped them to publishing year in figure 2.

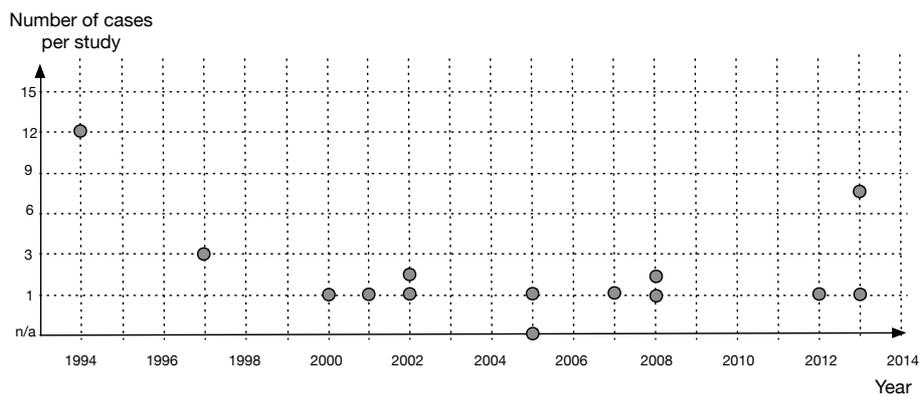

**Fig. 2.** Publishing years and number of cases per report

Table 3 summarizes the extracted publishing venues. A majority of the studies (60%) are published as conference papers.

**Table 3.** Publishing venues

| Publishing venue | Papers |
|---|---|
| IEEE Software | [1, 7, 30] |
| XP Conference | [29] |
| HCI International Conference | [35] |
| Lean Enterprise Software and Systems | [4] |
| International Journal of Project Management | [13] |
| International Conference on eXtreme Programming and Agile Processes in Software Engineering | [14] |
| Canadian Society for the Study of Education conference | [19] |
| Pacific Northwest Software Quality Conference | [18] |
| Agile conference | [21] |
| IEEE Computer | [36] |
| Americas Conference on Information Systems | [27] |
| SOFTWARE PROCESS—Improvement and Practice | [20] |

## 5 Analysis and Discussion

### 5.1 RQ1: What is state-of-practice in terms of utilization of software engineering knowledge areas in startups?

The mapping of practices to SWEBOK (table 2) shows that the majority of the main knowledge areas (11 out of 15) are addressed. However, a more detailed analysis reveals that only 28 of 62 categories from the knowledge areas are covered. One could argue that some of the knowledge areas, for example Mathematical Foundations knowledge area (KA), may be of less interest for startups or some categories could be more relevant than others. To better understand which knowledge areas and categories are more relevant for startups, we use Crowne's model of the startup life cycle [11].

We use Crowne's startup life-cycle model, in combination with the knowledge areas proposed by SWEBOK [31], to analyze whether the state-of-practice addresses software engineering challenges relevant for startups and to what extent such support is still lacking.

During the *startup* phase in Crowne's model, a company aims to build the first version of a product [11]. Understanding and communicating the needs of the target audience, and defining a development scope establish the foundation for further software engineering. The Requirements Engineering KA aims to support activities related to understanding needs and constraints placed on a software product, and is addressed by [1, 14, 19, 20, 29, 35]. Identified knowledge areas cover all categories, except Software Requirements Fundamentals and Software Requirements Tools. The Software Requirements Fundamentals category provides underlying concepts for the whole KA. For example, in this category the differentiation between functional and quality requirements is introduced. May [21] argues that a key differentiator between competitor products is an interaction experience, however the presence of specific quality requirements was not reported in his study. We argue that a lack of research in this area indicates an insufficient understanding of quality requirements' role in software engineering in the startup context.

Operating with very limited resources, a startup must carefully select the scope of the first release. Both scope definition and assessment belong to the SWEBOK Software Engineering Management KA, which is not addressed by any of identified studies. We argue that the absence of practices addressing scope definition could be a contributing factor to premature failure.

Following the *startup* phase, the *stabilization* phase [11] aims at improving the product to a level where it can be decommissioned to any number of new customers without causing any overhead on product development. The Software Design KA provides support for improving internal qualities of the product and is addressed by [1, 14, 18, 21, 29, 30, 35]. The Requirements Management category becomes relevant to maintain product integrity while adding new features [11], however this category is not addressed by any of identified studies.

After the *startup* and *stabilization* phases, the *growth* phase poses challenges like expanding the team, ensuring transfer of know-how, and managing the product. The Communication Skills category, addressing knowledge transfer within the team, is covered by [1, 19, 21]. The Product Life Cycle and Portfolio Management categories belong to the Software Engineering Economics KA, however none of the identified practices address these categories. The Software Engineering Economics KA directly addresses the relation between software technical decisions and business goals of the organization. We argue that absence of practices belonging to this area reveals a key gap in building viable software products in startups.

The *maturity* phase is the final phase on Crowne's model and it takes place when product development is robust and processes are predictable for day-to-day operations and invention of new products [11]. The Software Engineering Process KA addresses process introduction and improvement. Practices belonging to Software Process Measurement Techniques and Software Engineering Process Tools categories are reported in [1, 20]. Other categories of this KA are not covered by any of the identified practices. We argue that at this phase, startups gradually mature towards small-medium enterprises (SME), rendering research on software process introduction and improvement in SME's also relevant.

## 5.2 RQ2: What is the relevance and rigor of the studies reporting experiences from software engineering in startups?

Studies conducted in a realistic environment, e.g. a startup company, have a larger potential to provide useful results, compared to laboratory experiments [17]. A research method that facilitates investigation in realistic contexts, with industry professionals and on a realistic scale, contributes to industry relevance [17]. Moreover, the extent to which a study method is described contributes to the understanding of results and the evaluation of potential benefits and risks prior to application [17]. The rigor of the evaluation and presentation is also an indication to a level of trust that can be put on the results [17].

We have found that most identified studies are conducted in collaboration with actual startup companies, thus scoring high on relevance scale (figure 1). However, research type analysis suggests that most papers are experience reports (figure 1) and study only one case (figure 2). Further analysis shows that most of the papers fall into the low rigor category (figure 1). This implies that a) a majority of the studies do not compare and analyze data from multiple cases and b) results among different studies are difficult to compare due to their low rigor. Therefore, the extent to which reported results can be generalized is low, and transfer to different startup companies is difficult.

# 6 Conclusions

We have mapped software engineering practices from peer-reviewed scientific papers about startups to SWEBOK categories and to startup life cycle phases. This was done in order to understand to what extent software engineering in startups is supported by research. Results show that a surprisingly small number of papers address the core software engineering knowledge areas in startups. Even though this gap was first identified by Sutton et al. [33] more than a decade ago, very little has been done to address it.

By means of a literature review we have identified 54 practices that, to some extent, cover all critical knowledge areas. However, a majority of categories are not addressed by research. We analyzed whether the reported practices are actually useful for startups. Even though many knowledge areas are covered, we identified gaps in practices supporting successful transition trough the startup life cycle, particularly in market-driven requirements engineering, engineering scope definition, alignment between technical decisions and business goals, software architecture, and implementation of software engineering process.

The analysis of transferability of practices shows that the majority of studies are conducted in a realistic environment, thus providing relevant results. However the rigor of identified studies is low due to insufficient descriptions of applied research methods and poorly reported study contexts. In such an applied field as software engineering, the ability to transfer results from one environment to another is critical [17]. As a result, a lack of rigor makes this transfer difficult or even dangerous for two reasons. First, contextual information enables a company to see if a good practice or lesson reported is relevant in their context. Second, as study design details are missing the level of trust in how the study was performed is hard to judge. This result confirms similar conclusions by Paternoster et al. [23].

We conclude that existing studies, addressing software engineering in startups, are insufficient to support all engineering aspects and do not create a solid body of knowledge. Moreover, results from existing studies are hard to transfer to startup companies due to an inadequate level of reporting rigor.

While the mapping of engineering practices presented in this paper can serve as a basis, more empirical research with focus on product engineering in the start-up context is required to address the identified gap. Even though performing research in startups is difficult due to rapidly changing environment, more primary studies are needed to understand how software-intensive product engineering is performed in startups. Completing the picture on what practices are actually used in startups and what impact said practices had on product engineering process would be a first step. Identifying inadequacies in used practices and proposing remedies are our mid-term goals.